\documentclass{kapedbk}
\usepackage{graphicx,amsmath,amsfonts,edbkps}
\normallatexbib
\begin{document}
\articletitle{On the Electron-Electron Interactions in Two
Dimensions}
\chaptitlerunninghead{Electron Interactions in 2D}
\author{V.~M.~Pudalov}
\affil{P.~N.~Lebedev Physics Institute, Moscow 119991, Russia}
\email{pudalov@mail1.lebedev.ru}
\author{M.~E.~Gershenson}
\affil{Physics and Astronomy, Rutgers University, Piscataway, NJ
08854, USA}
\email{gersh@physics.rutgers.edu}
\and
\author{H.~Kojima}
\affil{Physics and Astronomy, Rutgers University, Piscataway, NJ
08854, USA}
\email{kojima@physics.rutgers.edu}

\begin{abstract}
In this paper, we analyze several experiments that address the
effects of electron-electron interactions in 2D electron (hole)
systems in the regime of low carrier density. The interaction
effects result in renormalization of the effective spin
susceptibility, effective mass, and $g^*$-factor. We found a good
agreement among the data obtained for different 2D electron
systems by several experimental teams using different measuring
techniques. We conclude that the renormalization is not strongly
affected by the material or sample-dependent parameters such as
the potential well width, disorder (the carrier mobility), and the
bare (band) mass. We demonstrate that the apparent disagreement
between the reported results on various 2D electron systems
originates mainly from different interpretations of similar ``raw''
data. Several important issues should be taken into account in the
data processing, among them the dependences of the effective mass
and spin susceptibility on the in-plane field, and the temperature
dependence of the Dingle temperature. The remaining disagreement
between the data for various 2D electron systems, on one hand, and
the 2D hole system in GaAs, on the other hand, may indicate more
complex character of electron-electron interactions in the latter
system.

\end{abstract}


\section{Introduction}
Understanding the properties of strongly interacting and
disordered two-dimensional (2D) electron systems represents an
outstanding problem of modern condensed matter physics. The
apparent "2D metal-insulator transition" (2D MIT) is one of the
puzzling phenomena that are still waiting for an adequate
theoretical description \cite{krav94,krav95}. Figure 1 shows that
the transition from the "metallic" to "insulating" behavior occurs
as the density of electrons $n$ is decreased below a certain
critical value $n_c$. The strength of electron-electron ($e$-$e$)
interactions is characterized by the ratio of the Coulomb
interaction energy to the  Fermi energy. This ratio, $r_s$,
increases $\propto 1/n^{1/2}$ \cite{ando_review} and reaches $\sim
10$ at $n\approx n_c$; this suggests that the $e$-$e$ interactions
might be  one of the major driving forces in the phenomenon. Thus,
better understanding of the properties of 2D systems at low
densities, and, in particular, in the critical regime \cite{akk}
in the vicinity of the apparent 2D MIT, requires quantitative
characterization of electron-electron interactions.

\begin{figure}[ht]
\vskip.2in
\begin{center}
\includegraphics[width=.8\textwidth]{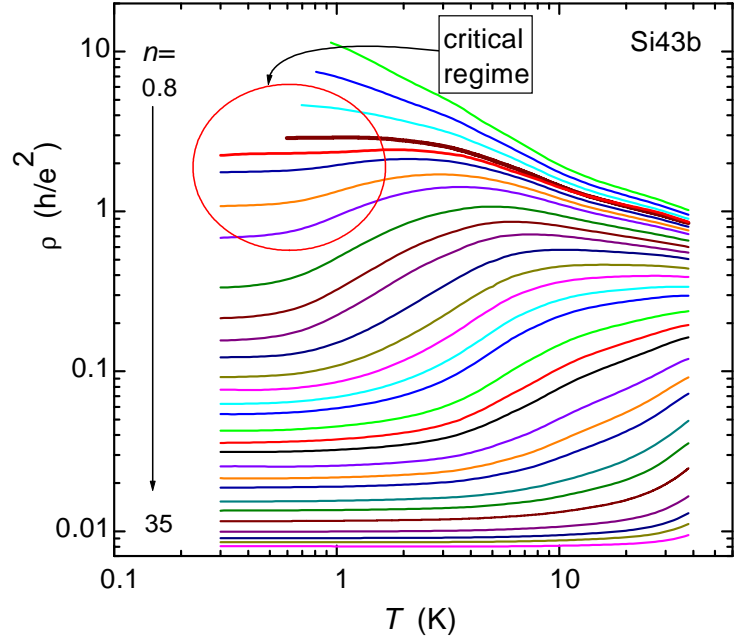}
\caption{Temperature dependences of the resistivity for Si-MOS device
over a  wide density range, 0.8 to $35\times 10^{11}$cm$^{-2}$
\protect\cite{akk,weakloc}.}
\end{center}
\label{fig:Si43_Domains}
\end{figure}

Within the framework of Fermi-liquid theory, the interactions lead
to renormalization of the effective quasiparticle parameters, such
as the spin susceptibility $\chi^*$, effective mass $m^*$,
Land\'{e} factor $g^*$, and compressibility $\kappa^*$.
Measurements of these renormalized parameters are the main source
of experimental information on interactions. The renormalizations
are described by   harmonics of the Fermi-liquid interaction in
the singlet (symmetric, (s)) and triplet (antisymmetric, (a))
channels, the first of them being:
\begin{equation}
F_0^a = \frac{2}{g^*}-1, \qquad F_1^s= 2\left(\frac{m^*}{m_b} -1
\right).
\end{equation}
Here $g_b$ and $m_b$ are the band values of the $g$-factor and
mass, respectively.

Recently, as a result of extensive experimental efforts, rich
information on the renormalized quasiparticle parameters has
become available for 2D systems. The corresponding results were
obtained by different techniques and for different material
systems. At first sight, the data sets in different publications
seem to differ from each other a great deal. Our goal is to
review briefly the available data and to analyze the sources of
their diversity. We find that, in fact, the apparent diversity
between various results originates mainly from different
interpretation of {\it similar} "raw" data. Being treated on the
same footing, most experimental data do agree with each other.
The remaining disagreement between the data for $p$-type GaAs, on
one hand, and the other systems, on the other hand, may indicate
more complex character of interactions in the former 2D hole
system.

\section{Renormalized spin susceptibility}
Several experimental techniques have been used for measuring the
renormalized spin susceptibility $\chi^*$, such as\\
(i) analysis
of the beating pattern of Shubnikov-de Haas (SdH) oscillations in
weak tilted or crossed magnetic fields
\cite{okamoto,gm},\cite{crossed,zhu};\\
(ii) fitting the temperature- and
magnetic field dependences of the resistivity
\cite{savchenko_RT,shashkin_RT},\cite{vitkalov_RB,aleiner} with the
quantum corrections theory \cite{ZNA,Gornyi};\\
(iii) the
magnetoresistance scaling in strong fields
\cite{vitkalov_scalingMR},\cite{shashkin_scalingMR},\cite{vitkalov_PRB};\\
(iv)
measuring the ``saturation'' or
hump in magnetoresistance in strong in-plane fields
\cite{yoon_2000},\cite{tutuc_2001},\cite{tutuc_2003},\cite{tutuc_2002},\cite{vitkalov_PRB,aniso};\\
(v) measuring the
thermodynamic magnetization \cite{reznikov}. \\ We compare below
the available experimental results.

\begin{figure}[ht]
\vskip.2in
\begin{center}
\includegraphics[width=.6\textwidth]{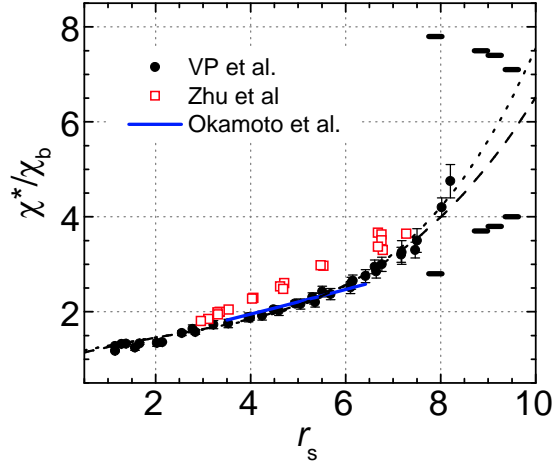}
\caption{Renormalized spin susceptibility measured by SdH effect
in tilted or crossed fields on n-SiMOS by Okamoto et al.
\protect\cite{okamoto}, Pudalov et al \protect\cite{gm}, and on
n-GaAs/AlGaAs by Zhu et al. \protect\cite{zhu}. Horizontal bars
depict the upper and lower limits on the $\chi^*$ values,
determined from the sign of SdH oscillations, measured at
$T=0.027$mK for sample Si5 \protect\cite{polarization}. Dashed and
dotted lines show two examples of interpolation of the data
\protect\cite{okamoto,gm}.} \label{fig:chi}
\end{center}
\end{figure}

\vspace{0.1in}
\underline{(1) SdH oscillations: $n$-Si and
$n$-GaAs.}
\\ Figure \ref{fig:chi} shows the $\chi^*(r_s)$ data obtained
by Okamoto et al. \cite{okamoto} for $n$-(100)Si-MOS system by
observing how the first harmonic of SdH oscillations vanishes in
tilted magnetic fields (the so called ``spin-zero'' condition,
which corresponds to the equality $g^*\mu_B B_{\rm tot}
=\hbar\omega_c/2$, where $B_{\rm tot}=\sqrt{B_\perp^2 +
B_\parallel^2}$ and $\mu_B$ is the Bohr magneton). More recent
results \cite{gm} on $n$-(100)SiMOS samples have been obtained
from the SdH interference pattern in weak crossed magnetic fields
\cite{crossed}; they extend the earlier data to both higher and
lower $r_s$ values. It is worth noting that the data presented in
Fig.~\ref{fig:chi} have been obtained for many Si-MOS samples
fabricated by different manufacturers \cite{gm,okamoto}; the peak
mobilities for these samples range by a factor of $\sim 2$.
Nevertheless, there is  a good agreement between the data for
different samples. We conclude therefore that {\em the effect of
disorder on the renormalization of $\chi^*$ at $n>n_c$ is
negligible or, at least, weak}.

As seen from Fig.~\ref{fig:chi}, the data on $n$-channel Si-MOS
samples are in a reasonable agreement with the data obtained by
Zhu et al. \cite{zhu} for $n$-type GaAs/AlGaAs samples using a
similar technique (measuring SdH effect in tilted magnetic
fields). Because of a smaller (by a factor of 3) electron
effective mass in GaAs, similar $r_s$ values have been realized
for the electron density 10 times lower than in Si-MOS samples.
The width of the confining potential well in such GaAs/AlGaAs
heterojunctions is greater by a factor of 6 than in (100) Si-MOS,
due to a smaller mass $m_z$, lower electron density, and higher
dielectric constant. This significant difference in the thickness
of 2D layers may be one of the reasons for the  20\% difference
between the $\chi^*$-data in $n$-GaAs and $n$-SiMOS samples seen
in Fig.~\ref{fig:chi}; at the same time, the minor difference
indicates that {\em the effect of the width of the potential well
on renormalization of $\chi^*$ is not strong.}

\begin{figure}[ht]
\vskip.2in
\begin{center}
\includegraphics[width=.8\textwidth]{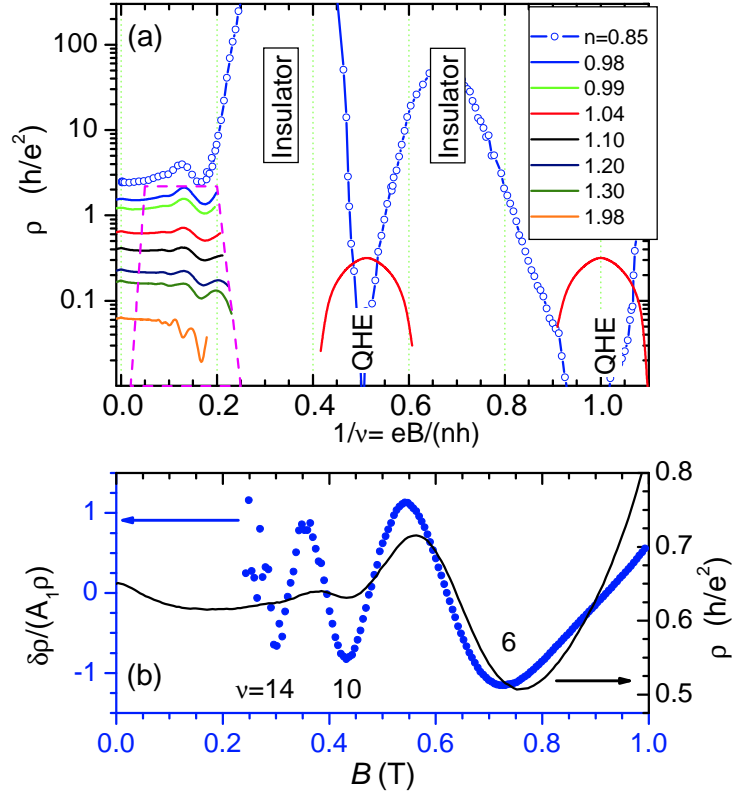}
\caption{(a) Overall view of the SdH oscillations in low fields at
different densities. Empty circles show the $\rho_{xx}$
oscillations for sample Si9 in high fields, corresponding to the
reentrant QHE-insulator transitions \protect\cite{PRB92}. (b)
Expanded view of one of the $\rho_{xx}(B)$ curves ($n=1.04\times
10^{11}$cm$^{-2}$ (right axis) and its oscillatory component
normalized by the amplitude of the first harmonic $A_1(B)$ (left
axis) \protect\cite{gm}.
Dashed line confines the region of the
SdH measurements in Refs.~\protect\cite{gm,polarization}.}
\end{center}
\label{fig:SdH_Domain}
\end{figure}

The SdH experiments provide the direct measurement of $\chi^*$ in
weak perpendicular and in-plane magnetic fields $\hbar\omega_c \ll
E_F$, $g^*\mu_B B_{\rm tot} \ll E_F$ \cite{gm,crossed}. Under such
conditions, the quantum oscillations of the Fermi energy may be
neglected, and the magnetization remains a linear function of $B$,
\, $\chi^*(B_{\rm tot})\approx \chi^*_0$. Also, under such
experimental conditions, the filling factor is large, $\nu
=(nh)/(eB_\perp) \gg 1$ and the amplitude of oscillations is small
$|\delta\rho_{xx}|/\rho_{xx} \ll 1$. Figure~\ref{fig:SdH_Domain}
shows, on the $\rho - B_\perp $ plane, the domain of the weak
magnetic fields, $\nu >6$, where the SdH oscillations have been
measured in Refs.~\cite{gm,polarization}. As the perpendicular
magnetic field increases further (and $\nu$ decreases), the SdH
oscillations at high density $n\gg n_c$ transform into the quantum
Hall effect; for low densities, $n\approx n_c$, the SdH
oscillations transform into the so-called ``reentrant
QHE-insulator''(QHE-I)  transitions  \cite{dio90,PRB92}. The
uppermost curve (open circles) presents the $\rho(B)$ variations
in the regime of QHE-I transitions \cite{dio90,PRB92}), measured
for a density slightly larger (by 4\%) than the critical value
$n_c$. This diagram is only qualitative, because the $n_c$ value
is sample-dependent.

\underline{Regime of low densities.} In the vicinity of the
critical density $n\approx n_c$, the number of observed
oscillations decreases,  their period increases, and the
interpretation of the interference pattern becomes more difficult,
thus  limiting the range of direct measurements of $\chi^*(r_s)$.

The horizontal bars in Fig.~\ref{fig:chi} are obtained from
consideration of the sign and period of SdH oscillations
\cite{polarization} as explained below. They show the upper and
lower limits for $\chi^*$, calculated from the data reported in
Refs.~\cite{gm,PRB92,polarization}. Figure~\ref{fig:SdH_Domain}\,b
demonstrates that in the density range $0.7<n<1\times
10^{11}$cm$^{-2}$, the oscillatory $\rho_{xx}$ (beyond the
magnetic field enhanced $\nu=1$ valley gap) has minima at filling
factors
\begin{equation}
\nu =(4i - 2), \qquad i= 1, 2, 3...,
\label{eq:minima_low_n}
\end{equation}
rather than at $\nu= 4i$ (in (100) Si-MOSFETs, the valley degeneracy
$g_v=2$). The latter situation is typical for high densities
and corresponds to inequality $g^*\mu_B B < \hbar\omega^*_c/2$.

In other words, the sign of oscillations at low densities is
reversed. This fact is fully consistent with other observations
(see, e.g., Fig.~2 of Ref.~\cite{polarization}, Fig.~1 of
Ref.~\cite{termination}, and Figs.~1-3 of
Ref.~\cite{krav_SdH_low_fileds}). As figure~\ref{fig:chi} shows,
the ratio $\chi^*/\chi_b$ exceeds $1/2m_b=2.6$ at $r_s \approx 6
$; the first harmonic of oscillations disappears at this density
(the so-called ``spin-zero''), and the oscillations change sign
for lower densities. Since the sign of the SdH oscillations is
determined by the ratio of the Zeeman to cyclotron splitting
\cite{LK,bychkov_SdH}
\begin{equation}
\cos\left(\pi \frac{g^*\mu_B B}{\hbar\omega^*_c}\right) \equiv
\cos\left(\pi \frac{\chi^*}{\chi_b}m_b\right),
\end{equation}
it was  concluded in Ref.~\cite{polarization} that, in order to
have negative sign in the range $10>r_s>6$, the spin
susceptibility
$\chi^*$ must obey  the following inequality:
\begin{equation}
2.6 = \frac{1}{2m_b} <\frac{\chi^*}{\chi_b} <\frac{3}{2m_b} = 7.9.
\label{eq:chi_low_n}
\end{equation}

Thus, Eq.~(\ref{eq:minima_low_n}) and Eq.~(\ref{eq:chi_low_n}))
enable us to set the upper and lower limits for $\chi^*$
\cite{polarization}, which are shown by horizontal bars in
Fig.~\ref{fig:chi} at $r_s =7.9-9.5$. As density decreases (and
$r_s$ increases), due to finite perpendicular fields, in which the
SdH oscillations were measured, the condition
Eq.~(\ref{eq:chi_low_n}) becomes more restrictive, which leads to
narrowing the interval between the upper and lower bars
 \cite{polarization}.

\underline{(2) Magnetoresistance in the in-plane field.}\\
Monotonic magnetoresistance (MR) in the in-plane field exhibits a
well-defined saturation for the $n$-type Si MOSFETs
\cite{bishop,simonian},\cite{JETPL_98b},\cite{breakdown,popovic_2002},
\cite{vitkalov_doubling_PRL},\cite{vitkalov_scalingMR}
or a hump for  the $n$- or $p$-type 2D GaAs systems
\cite{yoon_2000,tutuc_2002},\cite{tutuc_2001,tutuc_2003},\cite{savchenko_RT,zhu}. With
increasing mobility (and corresponding decreasing critical $n_c$
density), the latter hump  becomes more pronounced; it resembles
the sharp transition to the $R(B_\parallel)$ saturation in Si-MOS
\cite{noh_0206519}.

The hump or saturation of the
in-plane magnetoresistance have been
interpreted in Refs.~\cite{tutuc_2001,vitkalov_PRB} as a signature
of complete spin polarization $B_{\rm pol}$.
This treatment is
also supported by the experiments by Vitkalov et al.
\cite{vitkalov_doubling_PRL,vitkalov_doubling_PRB}
and Tutuc et al. \cite{tutuc_2003}, who found that
the frequency doubling of SdH oscillations  coincides with the
onset of saturation of the in-plane magnetoresistance.
Another
approach to the high field  measurements of $\chi^*$ is based on
the scaling of $R(B_\parallel)$ data
\cite{shashkin_scalingMR,vitkalov_scalingMR}: by scaling, the
$R(B_\parallel)$ data  for different densities are forced to
collapse onto each other. This procedure is essentially the
high-field one, $g^*\mu B \sim 0.6E_F$, as the chosen scaling
field $ B_{\rm sc} \approx 0.3B_{\rm pol}$.

The features in $\rho(B_\parallel)$ are observed  at a field
$B_{\rm sat}$, which is close to the estimated field of the
complete spin polarization \cite{breakdown}:
\begin{equation}
B_{\rm sat} \approx B_{\rm pol}= 2E_F/g^*\mu_B.
\label{eq:pol&sat}
\end{equation}
By assuming that $B_{\rm pol} = B_{\rm sat}$ and using the
standard expression for the 2D density of states, ${\rm DOS} = m^*
g_v/\pi\hbar^2$, one can estimate $\chi^*$ from measurements of
the characteristic field $B_{\rm sat}$:
\begin{equation}
g^*m^* = \frac{2n\pi\hbar^2}{B_{\rm sat}g_v \mu_B}.
\label{eq:polarization}
\end{equation}

Evaluation of  $\chi^*$ from the
aforementioned experiments in strong fields and from
Eqs.~(\ref{eq:polarization}) and (\ref{eq:pol&sat})
is based on the following assumptions: (i) $\chi^* \propto g^*m^*$
is $B_\parallel $-independent; (ii) $m^*$ and 2D DOS are
energy-independent. In general, both assumptions are dubious.
Nevertheless, for some samples, Eqs.~(\ref{eq:polarization}) and
(\ref{eq:pol&sat}) may give plausible results over a limited range
of densities. For example, the low-field SdH data and the high
field magnetoresistance data were found to differ only by $\leq
12\%$ over the density range $(1-10)\times 10^{11}$cm$^{-2}$. More
detailed critical analysis of the in-plane MR data may be found in
Refs.~\cite{zhu,tutuc_2003,aniso},\cite{reznikov,disorder,cooldown}.

An interesting interpretation of the MR data has been suggested in
Ref.~\cite{shashkin_scalingMR}, where the $1/\chi^*(n)$ dependence
determined down to $n=1.08\times 10^{11}$cm$^{-2}$, was linearly
extrapolated to zero at $n \approx 0.85\times 10^{11}$cm$^{-2}$
and interpreted  as an indication of the ferromagnetic instability
at this density. Our data, obtained from the analysis of the
period and sign of SdH oscillations at lower densities
\cite{polarization}, do not support this interpretation:\\
(i) in
the whole  domain of densities and fields depicted in
Fig.~\ref{fig:SdH_Domain}, no doubling of the frequency of SdH
oscillations is observed, which proves that the 2D system remains
spin-unpolarized (see e.g., Fig.~\ref{fig:SdH_Domain});\\
(ii) the
sign of the SdH oscillations [see Eq.~(\ref{eq:minima_low_n}) and
the discussion above] enables us to estimate the upper limit on
$\chi^*$ (see bars in Fig.~\ref{fig:chi}) in the interval of
$r_s=8 - 9.5$, i.e. $n= (1.08 - 0.77) \times 10^{11}$cm$^{-2}$.
Note that the latter interval  includes critical $n_c$ values for
most of the high mobility Si-MOS samples, in particular, those
used in Ref.~\cite{shashkin_scalingMR}.

\begin{figure}[ht]
\vskip.2in
\begin{center}
\includegraphics[width=.8\textwidth]{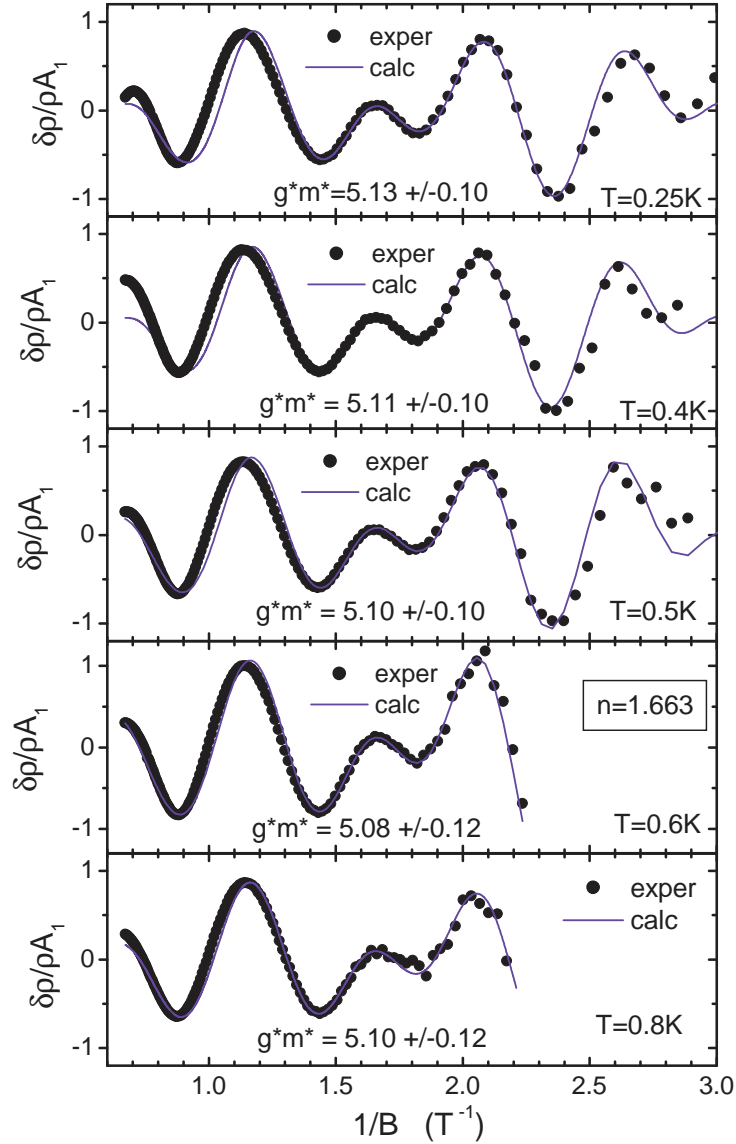}
\caption{Typical evolution of the interference pattern in SdH oscillations with
temperature. The oscillations are normalized by the
amplitude of the first harmonic \protect\cite{gm}.}
\end{center}
\label{fig:beating(T)}
\end{figure}

\underline{(3) Temperature dependence of $\chi^*$}.\\
In order to test
whether or not the
enhanced spin susceptibility $\chi^*$ depend
strongly on temperature,
we measured
the interference pattern of SdH oscillations for various
temperatures (see Fig.~\ref{fig:beating(T)}) and for different
densities, and thus determined the temperature dependence of
$\chi^*$. The results shown in Fig.~\ref{fig:beating(T)} reveal
only weak temperature variations of $\chi^*(T)$, within 2\% in the
studied $T$ range. We therefore can safely neglect the effect of
temperature in comparison of different sets of data.

There are several
possible reasons for the disagreement between the high-field
and low-field data; they are considered below.

\underline{(4) Effect of disorder on the high-field MR data.}\\
Firstly, it has been shown in Refs.~\cite{aniso,disorder} that the
saturation field $B_{\rm sat}$ and the high-field MR for
Si-MOSFETs \cite{cooldown} are strongly sample- (disorder-)
dependent. In particular, for a given density (and, hence, given
$E_F$), $B_{\rm sat}$ can vary by as much as a factor of two for
the samples with different mobilities. It was suggested in
Refs.~\cite{aniso,dolgopolov_comment},\cite{vitkalov_PRB} that these
variations are caused by the localized states, so that
Eq.~(\ref{eq:polarization}) might be thought to hold only for a
``disorder-free'' sample \cite{vitkalov_PRB}. However, by
extrapolating the measured $B_{\rm sat}$ fields
\cite{aniso,disorder} for samples with different peak mobilities
to $1/\mu^{\rm peak} \rightarrow 0$, one obtains a ``disorder
free'' $B_{\rm sat}^{\mu=\infty}$ value, which overshoots the spin
polarizing field \cite{disorder}, i.e. $B_{\rm sat}^{\mu=\infty} >
B_{\rm pol}$. This suggests that the structure of the localized
states below the Fermi level
is non-trivial
\cite{VP_reply to DG}.
Since $B_{\rm sat}$ crosses $B_{\rm pol}$, the two quantities
become equal at some mobility
value. For this curious reason, the
estimate Eq.~(\ref{eq:polarization}) provides correct results
\cite{krav_comment} for some samples with intermediate mobilities;
nevertheless, for lower densities $n\approx n_c$, deviations from
the SdH data are observed, as discussed in Ref.~\cite{VP_reply}.

\underline{(5) Magnetic field dependence of $\chi^*$}.\\
Secondly,
both parameters $m^*$ and $g^*$ (and  $\chi^* \propto g^*m^*$)
that enter Eqs.~(\ref{eq:pol&sat}), (\ref{eq:polarization}) depend on
the in-plane field. The $m^*(B_\parallel)$ dependence is mainly an
orbital  effect \cite{stern}; it is very strong for $n$-GaAs
samples with wider potential well \cite{zhu,tutuc_2003}. In contrast, the
$g^*(B_\parallel)$ dependence is apparently a spin-related effect
\cite{zhu,nonlinear}. The dependence of $m^*$ and $g^*$ on
$B_\parallel$ is another reason for the deviation of the
high-field $\chi^*$ values from the low field results  of SdH
measurements. In GaAs, the difference between the
low-$B_\parallel$ and high-$B_\parallel$ data is
dramatic \cite{zhu,tutuc_2003}: the density dependence of $\chi^*$ derived
for 2D electrons in GaAs  on the basis of the
$R(B_\parallel)$ measurements in high fields is non-monotonic \cite{zhu,tutuc_2003},
whereas the same samples demonstrated  a monotonic $\chi^*(n)$
dependence in low fields (see Fig.~\ref{fig:chi}) \cite{zhu,tutuc_2003}.
It is plausible, therefore, that ignoring the $m^*(B_\parallel)$
orbital dependence  causes the non-monotonic density dependence of
$F_0^a$, obtained in Ref.~\cite{noh_0206519} for 2D holes in GaAs
in the dilute regime $p\sim 10^{10}$cm$^{-2}$, in which the
potential well is very wide. The $m^*(B_\parallel)$ dependence is
also present in Si-MOS samples \cite{nonlinear}, though it is weaker than in GaAs
owing to a narrower potential well; as the density decreases and
potential well gets wider, this orbital effect should have a
stronger influence on the results of high-field MR measurements.

\begin{figure}[ht]
\vskip.2in
\begin{center}
\includegraphics[width=.65\textwidth]{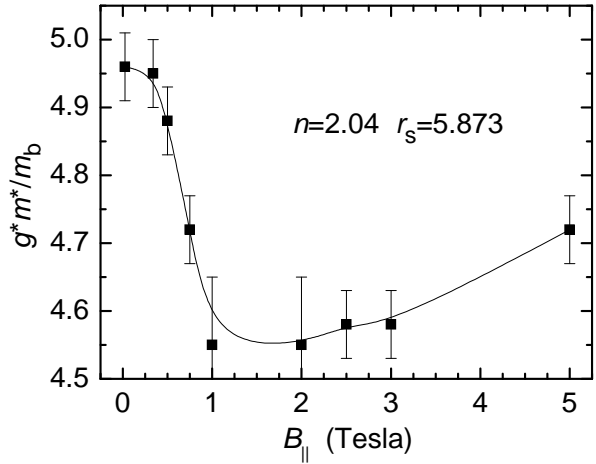}
\caption{Typical dependence of the spin susceptibility on the in-plane
magnetic field, measured for $n$-Si-MOS sample at $T\approx 0.15$K. Density
is given in units of $10^{11}$cm$^{-2}$.}
\end{center}
\label{fig:gm(B)}
\end{figure}

\underline{(6) Magnetization measurements.}\\ Another important
source of experimental information on the spin susceptibility are
the thermodynamic magnetization measurements, performed recently
by Reznikov et al. \cite{reznikov} on Si-MOS samples. Over the
density range $(3-9)\times 10^{11}$cm$^{-2}$, the measured $dM/dB$
is in agreement with the SdH data on $\chi^*$. The contribution of
localized states to the measured magnetization impedes the
detailed quantitative comparison with the SdH data at lower
densities. Nevertheless, two important results at low densities
are consistent with the SdH data: (i) the spin susceptibility
remains finite down to the lowest density (thus confirming the
absence of the spontaneous magnetization transition), and (ii) the
magnetization is nonlinear in $B_\parallel$ field with $\chi^*$
varying with field qualitatively similar to that shown in
Fig.~\ref{fig:gm(B)}.

\section{Effective mass and $g$-factor}
Historically, experimental data on the effective mass in 2D
systems have always been controversial (for a review of the
earlier data, see \cite{ando_review}). The data on $m^*$ have been
obtained mainly from the temperature dependence of SdH
oscillations. Even within the same approach, the data from
different experiments disagreed with each other at low densities.
With the advent of  high mobility samples, much lower densities
became accessible. However, the general trend remained the same:
disagreement between different sets of data grew as the density
was decreased; this disagreement becomes noticeable when $k_Fl$
becomes smaller than $\sim 5$.

\begin{figure}[ht]
\begin{center}
\includegraphics[width=.8\textwidth]{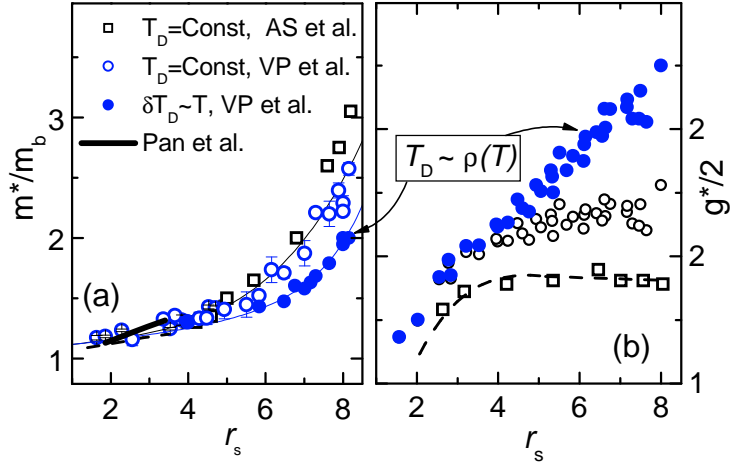}
\vspace{0.1in}
\begin{minipage}{3.2in}
\caption{Renormalized effective mass of electrons $m^*$ (a) and
renormalized $g$-factor (b) determined with Si-MOS samples in
different experiments as denoted in the legend. Data shown by open
boxes and circles are from Refs.~\protect\cite{shashkin_SdH}, and
\protect\cite{gm}, correspondingly, calculated using the LK
formula Eq.~(\protect\ref{eq:mass_FG}). Closed circles are the
data from Ref.~\protect\cite{gm} obtained using
Eq.~(\protect\ref{eq:mass_FL}).}
\end{minipage}
\end{center}
\label{fig:m(rs)}
\end{figure}

Figure~\ref{fig:m(rs)} shows that the data for Si-MOS samples
obtained in Refs.~\cite{gm,pan99},\cite{shashkin_SdH} are close to each
other only at $n>2.5\cdot10^{11}$ cm$^{-2}$ ($r_s <5$). At lower
densities, at first sight, there is a factor of $\sim 1.5$
disagreement between the data of Refs.~\cite{gm} and
\cite{shashkin_SdH} (closed and open symbols, respectively), which
is discouraging. However, we show below that the apparent
disagreement stems from different interpretations of raw data.
When treated on the same footing, the data agree reasonably well
with each other down to the lowest explored density
$n\approx1\times 10^{11}$ cm$^{-2}$ (i.e. $r_s \approx 8$).

One might suspect that the difference in the extracted $m^*$
values is due to the  different temperature ranges in different
experiments ($T=0.15-1$\,K and $0.3-3$\,K in Ref.~\cite{gm} and
$T=0.05-0.25$\,K in Ref.~\cite{shashkin_SdH}). However, the data
in Fig.~\ref{fig:beating(T)} do not reveal a strong $T$-dependence
of $\chi^*$. Since $\chi^*$ is proportional to $g^*m^*$, one has
to assume that the temperature dependences of $m^*$ and $g^*$ must
compensate each other; such compensation is highly unlikely.

In order to determine the effective mass from the temperature
dependence of the amplitude of SdH oscillations, one needs a
model; below we consider the models which are used in calculations
of $m^*$. The open squares \cite{shashkin_SdH} and open circles
\cite{gm} in Fig.~\ref{fig:m(rs)} are obtained by using the same
model of non-interacting Fermi gas, for which the amplitude of SdH
oscillations is given by the Lifshitz-Kosevich (LK) formula
\cite{LK}. The effective mass in this model is derived from the
$T$-dependence of the amplitude, which in the limit of $kT \gg
\hbar\omega_c$ can be expressed as:
\begin{equation}
-\frac{e\hbar H}{2\pi^2k_B c}\ln|\delta\rho_{xx}/\rho_{xx}| \approx
m^*(T+T_D).
\label{eq:mass_FG}
\end{equation}
If one assumes that the Dingle temperature  $T_D$ is {\em
temperature independent}, the calculated mass appears to depend on
the temperature interval of measurements \cite{gm}: the higher the
temperature, the larger the mass. Note that the direct
measurements of $g^*m^*(T)$ do not reveal any substantial
$T$-dependence of this quantity. Moreover, the mass value
calculated in this way was found to be somewhat different for
samples with different mobilities (i.e. $\tau$ values).

We believe that the aforementioned inconsistencies are caused by
assuming that $T_D$ is {\em temperature independent}. This
assumption is not justified, even if the resistance is
temperature-independent over the studied $T$ range (see, e.g.
\cite{maslov_SdH}). However, in a typical experimental situation,
determination of $m^*$ requires measurements of the oscillation
amplitude over a wide temperature range, where $\rho$ is strongly
$T$-dependent \cite{aleiner} owing to the interaction corrections
\cite{ZNA}.

In Ref.~\cite{gm}, in order to determine $m^*$ in a
strongly-interacting 2D electron system in Si MOSFETs, another
approach has been suggested, in which $T_D(T)$  was assumed to
reflect the temperature dependence of the resistivity
$\varrho=\varrho_0+\beta(n)T\tau$:

\begin{equation}
T_D^*(T) \approx
T_D(1+\beta(n) T\tau).
\label{eq:mass_FL}
\end{equation}
This empirical approach  eliminates largely the disagreement
between the results on $m^*$ for the same sample, obtained in
different temperature intervals, and between the results obtained
for different samples. This conjecture has been supported recently
by the theoretical study \cite{maslov_SdH}. The data shown in
Fig.~\ref{fig:m(rs)} by closed circles are obtained within this
approach \cite{gm};  we believe, they represent more reliable
$m^*$ data, which are consistent with the other types of
measurements (e.g., with the analysis  \cite{aleiner} of $\rho(T)$ in terms of the
theory of interaction corrections in the ballistic regime).

We will verify now whether or not the approach of
Eq.~(\ref{eq:mass_FL}) leads to convergence of the results from
Ref.~\cite{shashkin_SdH} (open boxes) and Ref.~\cite{gm} (closed
points). In order to do this, we use the $\sigma(T)$ dependences
reported in  Ref.~\cite{shashkin_RT} for the same samples. We
show below how the results on $m^*$ from Ref.~\cite{shashkin_SdH}
could be "corrected" in order to take into account a finite
$d\sigma/dT$). The open-box data point with the highest $r_s$
value in Fig.~\ref{fig:m(rs)} corresponds to the density
$n=1.03\times 10^{11}$cm$^{-2}$. The $\sigma(T)$ curves are
reported in Fig.~1 of Ref.~\cite{shashkin_RT} for two nearest
density values, $n=1.01$ and $1.08\times 10^{11}$cm$^{-2}$. For
simplicity, over the range of the SdH measurements $T=0.05-0.25$K
\cite{shashkin_SdH}, the $\sigma(T)$ dependence  may be
approximated by a linear $T$-dependence with the slope
$d\ln\sigma/dT \approx -1/$K.
According to Eq.~(\ref{eq:mass_FL}), we now use the $\sigma(T)$
slope together with
$T_D \approx 0.2$K  for the range $T=0.05-0.25$\,K as reported in
Ref.~\cite{shashkin_SdH}. As a result, we obtain
$T_{D}=T_{D0}(1+T)$ and, to the first approximation,
$m^*0.833[T+T_{D0}(1+T)] = m^*0.833[1.2 T+0.24]$ for the
temperature dependence of the logarithm of the oscillation
amplitude. The exact procedure of the non-linear data fitting
based on Eq.~(\ref{eq:mass_FL}) requires more thorough
consideration; we describe here a simplified step-by step
procedure of fitting. At the 2nd step one obtains  $0.8m^*[T 1.24
+0.248]$, etc. All the above functions fit equally well the same
raw data (i.e. the $T$-dependence of the  amplitude of
oscillations), but with different masses. Finally, the procedure
converges with the mass that is by $\sim 20 $\% smaller and the
Dingle temperature that is by 25\% larger than the initial values,
respectively. As a result, the disagreement between the data at
this $n$ in Fig.~\ref{fig:m(rs)} is reduced from 50\% to about
25\%.

We have repeated the same procedure at every density, for which
the $\sigma(T)$ curve is known for samples used in
Ref.~\cite{shashkin_SdH}. For the lower densities, similarity
between the results of Refs. \cite{gm} and \cite{shashkin_SdH} is
even more striking. For example, for the second data point
($n=1.08\times 10^{11}$cm$^{-2}$, $r_s=7.9$), the initial
disagreement between the masses is 43\%: $m^*/m_b=2.75$ (open
boxes) versus 1.92 (closed dots). After applying the same
procedure of the non-linear fitting with $d\ln\sigma /dT =
-0.72$K, and initial $T_D=0.25$K, we obtain the corrected values
$T_{D0}=0.333$K and $m^*/m_b=2.06$; the latter value differs only
by 7\% from our data (closed circles). At the highest density,
$n=2.4\times 10^{11}$cm$^{-2}$ ($r_s=5.37$), for which the
$\sigma(T)$ dependences are shown in Fig.~1 of
Ref.~\cite{shashkin_RT}, the mass correction is also
$\sim 6\%$.

Reduction of the $m^*$ values from Ref.~\cite{shashkin_SdH} (by
taking into account the $T$-dependence of $T_D$) leads to
re-evaluation of $g^*$: since $\chi^*(n)$ is known with higher
accuracy, the decrease in $m^*$ leads to the corresponding
increase in $g^* \propto \chi^*/m^*$. The $g^*(r_s)$ dependence
becomes monotonic, and comes into agreement with the earlier data
shown in Fig.~\ref{fig:m(rs)},b as closed dots.

\subsection{$F_0^a$ values}
\begin{figure}[ht]
\vskip.2in
\begin{center}
\includegraphics[width=.7\textwidth]{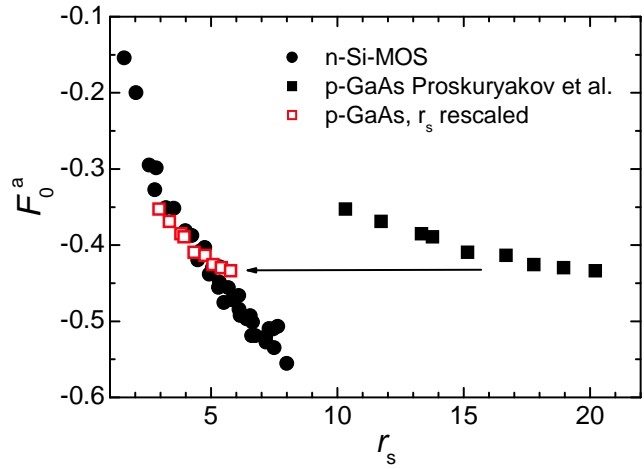}
\caption{Comparison of the $F_0^a$ values determined for n-SiMOS
\protect\cite{gm} and for $p$-GaAs/AlGaAs
\protect\cite{savchenko_RT}; the latter data are also shown versus $r_s$
without and with  scaling down by a factor of 3.5.}
\end{center}
\label{fig:F0(rs)}
\end{figure}

The $F_0^a$ values are determined from the renormalized $g^*$
factor,  Eq. (1.1). Firstly, as expected, we find that all
$g^*(r_s)$ data for (100) $n$-Si \cite{gm} and vicinal to (100)
Si-MOS samples \cite{vicinal_JPhysA} are rather close to each
other. Secondly, after the aforementioned correction has been
made to $m^*$, the data from Ref.~\cite{shashkin_SdH} become
consistent with the data from Ref.~\cite{gm}. Note that the
$m^*(r_s)$ and $F_0^a$ data for $n$-GaAs samples, determined on
the basis of approach Eq.~(\ref{eq:mass_FL}), are currently
unavailable. We focus below on comparison with $p$-GaAs, for
which the disagreement is dramatic, as Fig.~\ref{fig:F0(rs)}
shows.

\underline{Comparison with $p$-GaAs.}\\ With increasing quality of
$p$-GaAs/AlGaAs samples, the critical values of $r_s$ that
corresponds to the apparent 2D metal-insulator crossover grew from
17 \cite{savchenko_RT} to 37 \cite{noh_0206519}, and finally to
57 \cite{noh_0301301}. Observation of a non-insulating behavior at
such unprecedently high $r_s$ values represents a puzzle by
itself; two other puzzles are the observed non-monotonic behavior
of the renormalized $g$-factor (and $F_0^a$) with $r_s$
\cite{noh_0206519} and $r_s$-independent $m^*$
\cite{savchenko_RT}. Even if the nonmonotonic $g^*(r_s)$
dependence might be explained by the orbital effects (i.e. the
$m^*(B_\parallel)$ dependence) \cite{zhu}, the difference between
2D holes in GaAs and other 2D systems remains dramatic.

Clearly, the dependences $m^*(r_s)$ and $g^*(r_s)$ for $p$-GaAs
cannot be obtained by extrapolating the Si MOS data to higher
$r_s$ values (see Fig.~\ref{fig:m(rs)}). Not surprisingly,
therefore, that the $F_0^a$ data, deduced in
Refs.~\cite{savchenko_RT,proskuryakov_RB} from the temperature
dependence of the conductivity, differ substantially from the
values determined for $n$-Si-  and $n$-GaAs-based structures (see
Fig.~\ref{fig:F0(rs)}). It is highly unlikely that the values of
$F_0^a(r_s)$ ``jump up'' around $r_s\sim 10$ (where the data are
currently missing); such possibility is also at odds with the
numerical results. Rather, this non-monotonic dependence might
signal either the lack of the universal dependence $F_0^a(r_s)$ or
an incorrect quantification of the effective interaction strength
in different systems.

To choose between the aforementioned options, let us compare the
charge transport in $n$-type and $p$-GaAs systems in the
low-density regime. It is well-known that the experimental data
for various 2D electron and hole systems studied so far exhibit a
number of empirical similarities (quantitative within the same
host material and qualitative - for different systems). The two of
them are: (i) the relationship between the ``critical'' $\rho_c$
and $r_s$ values, and (ii) the magnitude of the resistivity drop
$\Delta\rho(T)/\rho_D$  at a given resistivity
$\rho_D$ value. Both
dependences imply a similar mechanism: the higher the quality of
the sample, the larger the critical $r_s$ (i.e. the lower $n_c$), $F_0^a$,
$\rho_c$ and the
magnitude of the resistance drop. These qualitative features have
been explained by the theory \cite{ZNA}, where the only
sample- (or disorder-) dependent parameter is the mean free time $\tau$ (the
higher  $\tau$, the stronger the ``metallic'' $\rho(T)$ dependence).

The low density $p$-GaAs
\cite{savchenko_RT,noh_0206519},\cite{noh_0301301} samples demonstrate
different features: on the one hand, the $r_s$-values are
extremely high (thus indicating a high sample quality and strong
interactions), on the other hand, the signatures of the metallic
behaviour are rather weak.  For the highest $r_s$ data
\cite{noh_0206519,noh_0301301} the renormalized Fermi energy is so
small ($\sim 0.1$K) that the 2D systems becomes non-degenerate
very quickly  as $T$ grows. This might explain the weak magnitude
of the resistance drop in the measurements of
Ref.~\cite{noh_0206519,noh_0301301}. However, this line of
reasoning cannot be applied to the higher-density (311) $p$-GaAs
samples \cite{savchenko_RT}, in which the Fermi energy is
larger. In order to bring  the above  data for $p$-GaAs into
agreement with other data, one has to scale the $r_s$ values down
by a factor of 6 \cite{savchenko_RT} and factor of 8
\cite{noh_0206519}.

It might be, therefore, that the effective $e$-$e$ interactions
are weaker in $p$-GaAs samples than in the other systems for the
same $r_s$ value, owing, e.g., to a more complicated physics of the
multivalley band structure and strong spin-orbit effects. If this
is the case, the interactions in $p$-GaAs samples cannot be
adequately quantified with a single parameter $r_s$. We illustrate
this in Fig.~\ref{fig:F0(rs)} by a simple rescaling of the
effective $r_s$ values for the data  on  $p$-GaAs
\cite{savchenko_RT}. Despite the raw data differ substantially,
they come into a reasonable agreement when $r_s$ for $p$-GaAs is
scaled down by an empirical  factor $3.5$. Of course, from this
rescaling, it is impossible to conclude whether the effective
$r_s$ values should be increased for $n$-Si- and $n$-GaAs- based
structures, or  decreased for $p$-GaAs; however, the multitude of
the material systems which show reasonably consistent data, points
at a somewhat  more complex behavior in $p$-GaAs. The same
empirical scaling procedure, being applied to the $m^*(r_s)$ data
for $p$-GaAs, helps to resolve another puzzle. The data for the
effective mass that were found in Ref.~\cite{savchenko_RT} to be
$r_s$ independent  over the range $r_s = 10 - 17$, after such
rescaling will fall into the range $r_s= 2.8 - 4.8$, where the
mass variations with $r_s$ are small (see Fig.~\ref{fig:m(rs)}).

\section{Summary}
To summarize, we compared various experimental data on the
renormalization of the effective spin susceptibility, effective
mass, and $g^*$-factor. If the data are considered on the same
footing, one finds a good agreement between different sets of
data, measured by different experimental teams using different
experimental techniques, and for different 2D electron systems.
The consistency of the data provides an additional evidence that
the renormalization is indeed caused by the Fermi-liquid effects.
The renormalization is not strongly affected by material- and
sample-dependent parameters such as the width of the potential
well, disorder (sample mobility) and the band mass value. The
apparent disagreement between the reported results is caused
mainly by different interpretation of similar raw data. Among  the
most important issues to be taken into account in the data
processing, there are the dependences of the effective mass and
spin susceptibility on the in-plane field, and the temperature
dependence of the Dingle temperature (the latter is intrinsic for
strongly-interacting systems). The remaining disagreement with the
data for 2D hole system in GaAs suggests that the character of the
effective electron-electron interaction is more complex in this
system; this important issue deserves thorough theoretical
attention.

\begin{acknowledgments}
The work was supported in part by NSF, ARO MURI, INTAS, RFBR, and
the Russian grants from the Ministry for Science and Technology,
Programs of the RAS, ``Integration of the high education and
academic research'',  and the Presidential Program of the support
of leading scientific schools.

\end{acknowledgments}

\end{document}